\documentclass[letterpaper]{article} 
\usepackage{aaai2026}  
\usepackage{times}  
\usepackage{helvet}  
\usepackage{courier}  
\usepackage[hyphens]{url}  
\usepackage{graphicx} 
\usepackage{natbib}  
\usepackage{caption} 
\usepackage{algorithm}
\usepackage{algorithmic}
\usepackage{booktabs}
\usepackage{amssymb}
\usepackage{amsmath}
\usepackage{makecell}

\usepackage{newfloat}
\usepackage{listings}
\DeclareCaptionStyle{ruled}{labelfont=normalfont,labelsep=colon,strut=off} 
\floatstyle{ruled}
\newfloat{listing}{tb}{lst}{}
\floatname{listing}{Listing}
\newcommand{\partitle}[1]{\smallskip \noindent \textbf{#1}.}
\usepackage[table]{xcolor}

\title{Walk Before You Dance: High-fidelity and Editable Dance Synthesis via Generative Masked Motion Prior}
\author{
    Foram Shah\equalcontrib, Parshwa Shah\equalcontrib, Muhammad Usama Saleem, Ekkasit Pinyoanuntapong, Pu Wang, Hongfei Xue, Ahmed Helmy
}
\affiliations{
    University of North Carolina at Charlotte\\
    \{fshah3, pshah77, msaleem2, epinyoan, pwang13, hxue2, ahelmy1\} @uncc.edu 
}

\begin{document}



\twocolumn[{
\renewcommand\twocolumn[1][]{#1}
\maketitle
\begin{center}
    \centering
    \includegraphics[width=0.9\textwidth]{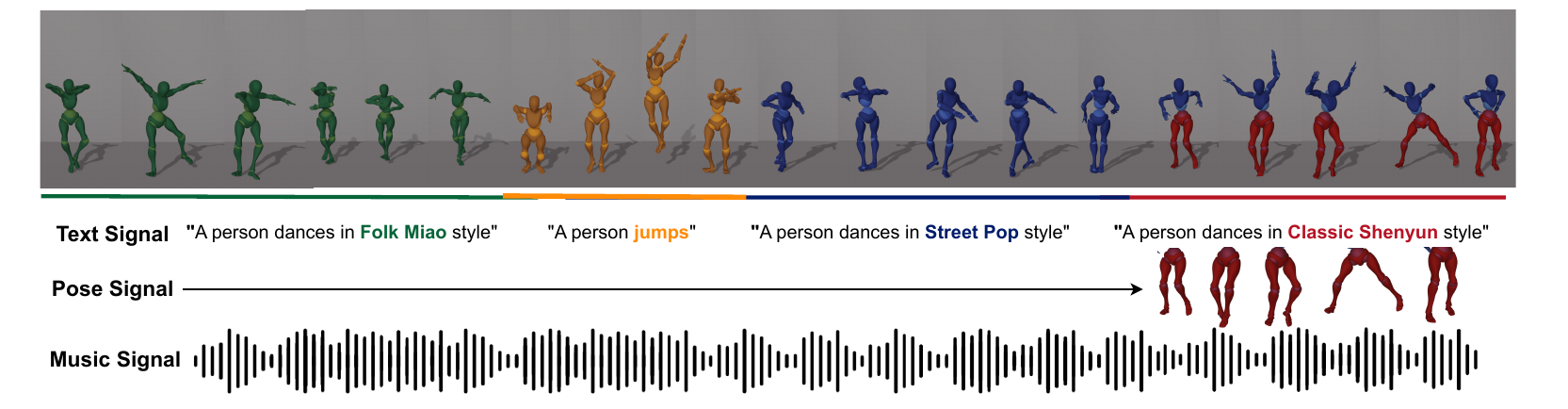}
    \captionof{figure}{DanceMosaic generates 3D dance motions based on multiple guidance signals. The top sequence showcases generated dance motions influenced by different text prompts, including genre-based or action-specific prompts. The color-coded figures represent different dance styles, synchronized with a music signal at the bottom. The pose signal allows further motion refinement, demonstrating the flexibility and precision of DanceMosaic.}
    \label{fig:teaser}
\end{center}
}]
\renewcommand{\thefootnote}{\fnsymbol{footnote}} 
\footnotetext[1]{Equal contribution} 

\begin{abstract}
Recent advances in dance generation have enabled the automatic synthesis of 3D dance motions. However, existing methods still face significant challenges in simultaneously achieving high realism, precise dance-music synchronization, diverse motion expression, and physical plausibility. To address these limitations, we propose a novel approach that leverages a generative masked text-to-motion model as a distribution prior to learn a probabilistic mapping from diverse guidance signals, including music, genre, and pose, into high-quality dance motion sequences. Our framework also supports semantic motion editing, such as motion inpainting and body part modification. Specifically, we introduce a multi-tower masked motion model that integrates a text-conditioned masked motion backbone with two parallel, modality-specific branches: a music-guidance tower and a pose-guidance tower. The model is trained using synchronized and progressive masked training, which allows effective infusion of the pretrained text-to-motion prior into the dance synthesis process while enabling each guidance branch to optimize independently through its own loss function, mitigating gradient interference. During inference, we introduce classifier-free logits guidance and pose-guided token optimization to strengthen the influence of music, genre, and pose signals. Extensive experiments demonstrate that our method sets a new state of the art in dance generation, significantly advancing both the quality and editability over existing approaches. 
\end{abstract}

\begin{links}
    \link{Project Page}{https://foram-s1.github.io/DanceMosaic/}
\end{links}

\section{Introduction}

Music-conditioned dance motion generation has recently garnered significant attention due to the intuitive rhythmic structure and semantic richness that music provides in guiding human motion. This field has broad applications in choreography, animation, virtual and augmented reality, and robotics. However, existing approaches \cite{zhuang2022music2dance,siyao2022bailando,li2023finedance, li2024lodge, tseng2023edge} still struggle to generate realistic, natural, and physically plausible 3D dance motions that not only reflect a specific dance genre, but also align precisely with musical cues such as rhythm, beat, and tempo. 

State-of-the-art dance generation models fall primarily into two categories: autoregressive and motion-space diffusion models. Autoregressive approaches, such as Bailando \cite{siyao2022bailando}, employ GPT-style next-token prediction to foster rhythmic consistency while preserving motion diversity. Nevertheless, their sequential nature often results in a misalignment between musical beats and the generated movements. In contrast, diffusion-based models like FineNet, LODGE, and EDGE \cite{li2023finedance, li2024lodge, tseng2023edge} achieve better synchronization between music and motion but often fail to produce high-fidelity dance sequences that match the realism and naturalness of authentic human movement. Beyond fidelity and alignment, editability is a critical feature for motion generation models. The capacity for editing tasks, such as modifying the upper or lower body or performing motion interpolation, is vital for iterative refinement and prototyping. This enables choreographers to fine-tune their concepts and supports the creative process in crafting original dances.

To overcome these challenges, we introduce DanceMosaic, a novel high-fidelity dance generation framework supporting a wide range of motion editing tasks guided by multiple input signals, including music, pose, and textual prompts describing actions or dance genres. Inspired by the human process of acquiring motor skills, first mastering foundational movements before progressing to advanced, expressive actions such as dance,  DanceMosaic’s central innovation lies in a multi-tower generative motion model combined with specialized training and inference strategies. Specifically, our model integrates a text-conditioned masked-motion backbone with two parallel modality-specific towers: a music-guidance tower and a pose-guidance tower. The backbone model first acquires basic motion skills from textual instructions, establishing a robust motion prior. Building upon this learned foundation, the music and pose-guidance towers further refine their capability to generate complex dance movements conditioned by musical cues and predefined pose constraints. To effectively achieve this hierarchical skill transfer, we employ synchronized and progressive masked training, enabling the motion priors learned by the text-to-motion model to be seamlessly infused into dance synthesis. This approach allows each guidance tower to optimize independently through its modality-specific loss, mitigating gradient conflicts. Additionally, during inference, we propose classifier-free logits guidance and pose-guided token optimization to further enhance the model's responsiveness to input signals, such as music rhythms, dance genres, and pose requirements.

\begin{itemize}
\item  We introduce DanceMosaic, the first method leveraging multi-tower generative masked models to enable high-fidelity dance generation with flexible editability.
\item  We propose synchronized and progressive masked training along with inference-time multimodal guidance modules, facilitating the synthesis of realistic, natural, physically plausible, and editable dance movements.
\item  Experimental results show that DanceMosaic establishes a new state-of-the-art performance in dance generation, significantly advancing the quality and editability achieved by existing approaches shown in Table \ref{tab:comparison}.
\end{itemize}

\begin{table}[t]
    \centering
    \scalebox{0.7}{
    \begin{tabular}{lccccc}
        \toprule
        Method & FID$_g\downarrow$ & Div$_g\uparrow$ & BAS $\uparrow$ & RunTime(s) $\downarrow$ & Editability \\
        \midrule
        FACT & 97.05 & 6.37 & 0.1831 & 9.46 & $\times$ \\
        MNET & 90.31 & 6.14 & 0.1864 & 10.26 & $\times$ \\
        Bailando & 28.17 & 6.25 & 0.2029 & \color{blue}{\underline{1.46}} & $\times$ \\
        EDGE & 50.38 & \color{blue}{\underline{6.45}} & 0.2116 & 2.27 & \checkmark \\
        LODGE & \color{blue}{\underline{34.29}} & 5.64 & \color{red}{\textbf{0.2397}} & 8.16 & $\times$ \\
        \hline
       \rowcolor{gray!15}
\textbf{DanceMosaic (Ours)} & \color{red}{\textbf{12.96}} & \color{red}{\textbf{8.36}} & \color{blue}{\underline{\textbf{0.2190}}} & \color{red}{\textbf{0.8}} & \color{red}{\checkmark} \\
\bottomrule
    \end{tabular}
    }
    \caption{DanceMosaic outperforms SOTA methods in terms of dance quality (FID), diversity (Div), and Run Time, without sacrificing dance-music alignment (BAS), while allowing mutimodal editing based on music, text, and pose guidance signals. (Red: best. Blue: runner-up)}  
    \label{tab:comparison}
\end{table}

\section{Related Work}

\subsection{Text-conditioned Motion Synthesis}
Early methods relied on motion matching \cite{holden2020learned}, while generative models have since advanced text-conditioned motion synthesis \cite{li2020learning, holden2016deep, rempe2021humor, zhou2019continuity, petrovich2021action, guo2020action2motion, ahuja2019language2pose, ghosh2021synthesis, yamada2018paired, TEMOS, athanasiou2022teach, TM2T}. Diffusion-based approaches refine motion through structured denoising \cite{kong2023priority, FLAME, MotionDiffuse} but suffer from high training complexity and slow inference. Autoregressive models \cite{MotionGPT, BAMM} leverage causal transformers for improved realism and diversity but struggle with fine-grained control and temporal consistency. More recent frameworks like Momask \cite{momask} and MMM \cite{pinyoanuntapong2024mmm} adopt masked motion modeling for better motion quality and diversity. Nevertheless, the use of masked modeling in music-conditioned dance generation remains unexplored due to fundamental semantic differences between textual and musical modalities. Existing masked motion methods typically follow a single-tower architecture tailored for text-conditioned tasks. In contrast, our proposed DanceMosaic approach introduces a novel multi-tower masked model designed specifically to support high-quality, editable dance synthesis guided by diverse multimodal inputs, including music signals, pose constraints, and textual prompts.

\subsection{Music-Conditioned Dance Synthesis}
Music-driven motion generation requires precise temporal synchronization between movement and rhythm. Early methods used motion retrieval and rule-based approaches, yielding beat-aligned but repetitive dances. Autoregressive models like Bailando \cite{siyao2022bailando} use GPT-like prediction to enhance rhythmic consistency while maintaining diversity. However, their sequential nature often causes misalignment between beats and movements. Recent diffusion-based models such as FineNet, LODGE, and EDGE \cite{li2023finedance, li2024lodge, tseng2023edge} improve dance-music alignment but struggle with generating diverse, high-fidelity dance. Their iterative denoising also leads to slow inference, limiting real-time usability. DanceMosaic overcomes these limitations by enabling high-quality and editable dance generation with real-time inference. Unlike diffusion models, it employs bidirectional BERT-like modeling \cite{MaskGIT, pinyoanuntapong2024mmm, BAMM}, while supporting multimodal dance generation through a novel multi-tower conditioned masked transformer.

\begin{figure*}[t]
    \hspace{5mm}
    \centering
    \includegraphics[width=0.9\textwidth]{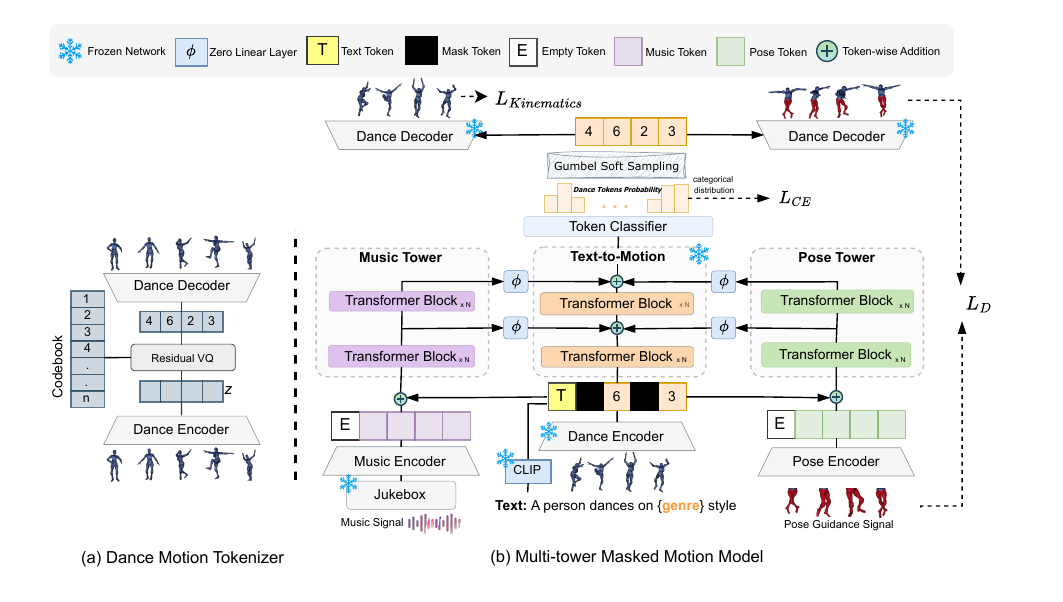}
\caption{Overview of DanceMosaic's training phase. (a) The process involves encoding dance motions into discrete token sequences using a dance motion tokenizer. (b) These tokens are then processed through a multi-tower masked motion model, where each tower (music, text, and pose) is used to learn the probabilistic mappings from modality-specific guidance signals to motion tokens. The model is trained using a progressive training strategy to integrate music, text, and pose signals.}
\label{fig:dancemosaic}
    \label{fig:main}
\end{figure*}

\section{Proposed Method: DanceMosaic}
\label{sec:method}
Given an input music signal $A$, a text prompt $T$, and pose constraints on specific joints or body parts $P$, our goal is to synthesize a physically plausible 3D dance sequence that aligns with the music and text prompts, while satisfying the pose constraints. To achieve this, we propose a multi-tower masked motion model, which incorporates music and pose guidance modules into the text-conditioned masked motion model. The overview of DanceMosaic, shown in Fig. \ref{fig:main}, consists of a dance motion tokenizer and three integrated masked transformers, including a Text-to-Motion (T2M) model, Music Tower (MT), and Pose Tower (PT). Motion tokenizer transforms the raw motion sequence into categorical discrete motion tokens within a learned codebook (Sec \ref{sec:tokenizer}). The T2M model learns to predict masked motion tokens, conditioned on textual prompts, like dance genres and action descriptions. The music and pose towers in Sec. \ref{sec:Towers} are designed to introduce the influence of music and pose signals to the T2M model via synchronized and progressive multimodal training (Sec. \ref{sec:training}). During inference, the multimodal guidance modules, including classifier-free logits guidance and inference-time token optimization (ITTO), steer the token sampling process so that generated dance sequence is aligned with diverse prompt signals (Sec. \ref{sec:inference_opt}). 

\subsection{Text-to-Motion Model as a Generative Prior}
\label{sec:tokenizer}
\textbf{Motion Tokenizer.} The goal of this module is to encode continuous dance motions into discrete categorical pose tokens using a learned codebook based on VQ-VAE \cite{vqgan}. This discretization is a critical step, as it enables subsequent generative masked models to learn the per-token categorical distribution conditioned on diverse input modalities. To unify the representation space, we train the motion tokenizer on a mixture of text-to-motion and music-to-motion datasets, enforcing a shared latent space across modalities. This joint training not only enhances the expressiveness of learned pose tokens but also allows us to leverage text-to-motion model as a strong motion prior for music-conditioned motion generation.

Given a motion sequence \( M = [m_1, m_2, \dots, m_N] \), where each frame \( m_i \in \mathbb{R}^D \) represents a 3D skeletal pose, where D is the dimension of human pose representations. The encoder compresses it into a latent representation \( Z \in \mathbb{R}^{n \times d} \) with a temporal downsampling factor of \( N/n \). The latent features \( Z = [z_1,z_2,...,z_{n}] \) are then quantized into discrete tokens \( \bar{Z} = [\bar{z}_1,\bar{z}_2,...,\bar{z}_{n}] \) from a learned codebook \( \mathcal{C} = \{c_l\}_{l=1}^{K} \), consisting of \( K \) unique code entries. The best-matching code is determined by minimizing the Euclidean distance \( \bar{z}_k = \operatorname{argmin}_{l} \| z_k - c_l \|_2^2 \). The tokenizer is trained using the following loss function:
\begin{equation}
\mathcal{L}_{\text{RVQ}} = \|M - \hat{M}\|_1 +  
\|\text{sg}(Z) - \bar{Z}\|_2^2 + \|Z - \text{sg}(\bar{Z})\|_2^2 
\end{equation}  
where \( \hat{M} \) is the reconstructed motion and \( \text{sg}(\cdot) \) represents the stop-gradient operation.  To minimize the quantization errors, we adopt Residual Vector Quantization (RVQ) \cite{momask}, which encodes a raw motion sequence into multiple token sequences within the latent space, with each sequence produced by a distinct quantizer. Each quantizer focuses on capturing the residual error from its predecessor.  More details can be found in the supplementary material. 


\begin{figure*}[t]
    \centering
    \includegraphics[width=\textwidth]{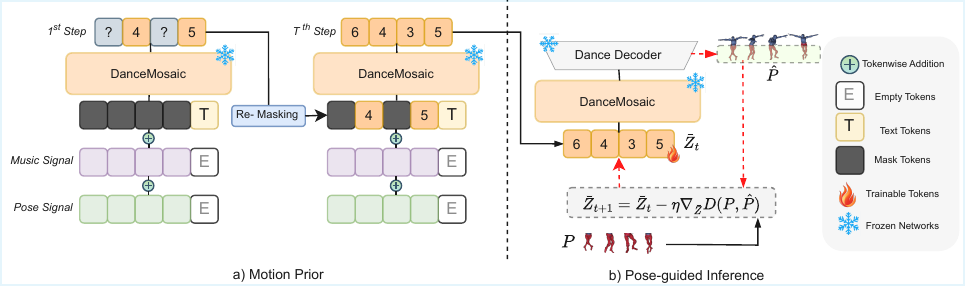}
    \caption{ Overview of DanceMosaic's inference phase. (a) Motion Prior: It parallelly encodes music, pose, and text conditions, passing the conditions through respective towers, which generates guidance for each modality. (b) Pose-guided Inference: At the final stage, we utilize inference time computing to refine the generated pose to align closely, if pose conditions are provided. }
    \label{fig:inferece_dancemosaic}
\end{figure*}

\noindent \textbf{Text-to-Motion Model (T2M).} Our T2M model employs a standard multilayer transformer, whose inputs are the concatenation of motion tokens $x_{1:t}$ from the tokenizer with $t$ as the sequence length, and the embedding $x_0$ from the pre-trained CLIP model \cite{CLIP} that takes both the dance prompt "A person dances on [genre] style" from dance dataset and the general text prompt from HumanML3D dataset. Due to the nature of self-attention in transformers, all motion tokens are learned in relation to the text embedding.  Given the discrete dance token sequence \( \bar{Z} = [\bar{z}_1,\bar{z}_2,...,\bar{z}_{n}] \), a subset of tokens is masked, forming a corrupted sequence \( Z_{\mathbf{M}} \). This sequence, along with a text signal \( T \), is processed by a text-conditioned masked motion transformer to recover the original dance motion. The model is trained to maximize the likelihood of correctly predicting masked tokens \cite{pinyoanuntapong2024mmm, momask}, i.e., minimizing the cross-entropy (CE) loss:
\begin{equation}
\mathcal{L}_{\text{CE}}^{T2M} = - \mathbb{E}_{\bar{Z}}  
\sum_{k \in \Omega}  
\log p_\theta (\bar{z}_k \mid Z_{\mathbf{M}}, T),
\end{equation}  
where \( \Omega \) denotes the set of masked indices and \( p_\theta (\bar{z}_k \mid Z_{\mathbf{M}}, T) \) is the parameterized probability of each motion token conditioned on $Z_{\mathbf{M}}$ and $T$. 



\subsection{Multi-tower Masked Motion Model}  
\textbf{Parallel Guidance Towers.} Building on the strong motion prior of the pre-trained T2M model, our model is designed to generate dance sequences conditioned on a music control signal A, a text prompt T, and pose guidance P. In particular, we integrate the text-conditioned masked motion model with two parallel modality-specific masked models, including a music-guidance tower and a pose-guidance tower. Each tower is implemented as a trainable counterpart to the original text-conditioned transformer. Each self-attention layer in the modality-specific tower is paired with a corresponding layer in the T2M model, connected via a zero-initialized linear layer ($\phi$). This design, inspired by diffusion ControlNet \cite{ControlNet}, ensures that the learned text-conditioned motion distribution does not interfere with music-to-dance mapping during early training. 

\noindent \textbf{Guidance Signals and Per-frame Alignment.}
\label{sec:Towers}
For the music tower, we adopt a frozen Jukebox model \cite{dhariwal2020jukebox} to provide a per-frame music embedding sequence, enabling precise synchronization between dance movements and musical rhythm. For pose tower, we introduce a spatial editing signal ${P} \in \mathbb{R}^{N \times J \times 3} $ into the masked token reconstruction, allowing targeted modifications to upper-body, lower-body or joint-specific movements. The pose signal ${P}$ encodes the $(x, y, z)$ positions of each of the $J$ skeleton joints over $N$ frames. A subset of joint locations is provided as spatial constraints and remains fixed during generation, while the positions of the remaining joints, which are set to zero, are editable. To effectively integrate these guidance signals, we leverage learnable music and pose encoders to transform the respective inputs into latent tokens matching the dimensionality of motion tokens. These music and pose latent tokens are then precisely aligned with motion tokens on a per-frame basis via token-wise addition, ensuring coherent synchronization across dance, music, and pose inputs.

\subsection{Synchronized Progressive Masked Training}  
\label{sec:training}
\textbf{Synchronized Masking.} The music and pose-guidance towers are trained using synchronized masked training, where both the T2M model and guidance towers receive identically masked motion token. This synchronized masking ensures that the guidance towers can effectively leverage the motion distributions learned by the T2M model. The music tower is trained to recover the masked motion tokens, conditioned on the text prompt \( T \), the music signal \( A \), and the corrupted motion token sequence $Z_{\mathbf{M}}$ by minimizing the CE loss:  
\begin{equation}
\mathcal{L}_{\text{CE}}^{MT} = - \mathbb{E}_{\bar{Z}}  
\sum_{k \in \Omega}  
\log p_\theta (\bar{z}_i \mid Z_{\mathbf{M}}, T, A),
\end{equation}  
where \( \Omega \) represents the masked indices. 

Besides CE loss  $\mathcal{L}_{\text{CE}}^{MT}$, we adopt additional kinematics losses, including joints position loss $\mathcal{L}_{\text{pos}}$, velocity loss $\mathcal{L}_{\text{vel}}$, acceleration loss in $\mathcal{L}_{\text{acc}}$, and foot loss $\mathcal{L}_{\text{foot}}$. These losses measure the difference in motion dynamics between the generated dance and ground-truth motion. Incorporating these losses can improve the physical plausibility and naturalness of diffusion-based dance generation \cite{tseng2023edge}.  However, integrating these kinematics losses into generative masked model training is difficult because it requires converting discrete pose tokens from the model's latent space into continuous Euclidean space.  This conversion requires sampling the categorical distribution of motion tokens during training, which is non-differentiable. To address this challenge, we employ a straight-through Gumbel-Softmax strategy \cite{li2020gumbelsoftmaxbasedoptimizationsimplegeneral} to make token sampling process differentiable by approximating discrete categorical distribution with continuous Gumbel-Softmax distribution.  The total loss function $\mathcal{L}_{\text{MT}}$ is the weighted sum of the CE loss $\mathcal{L}_{\text{CE}}^{MT}$ and kinematics losses ($\mathcal{L}_{Kinematics}$), i.e., \( \mathcal{L}_{MT} = \mathcal{L}_{\text{CE}}^{MT} + \lambda_{pos}\mathcal{L}_{\text{pos}} + \lambda_{vel}\mathcal{L}_{\text{vel}} + \lambda_{acc}\mathcal{L}_{\text{acc}} + \lambda_{foot}\mathcal{L}_{\text{foot}}  \). The details of $\mathcal{L}_{\text{MT}}$ are listed in supplementary materials.

The pose tower is trained to learn the motion distribution $p_{\theta}(\bar{z}_i \mid Z_{{\mathbf{M}}}, T, P)$  conditioned on corrupted motion tokens $Z_{\mathbf{M}}$, text prompt $T$, and pose signal $P$. 
\begin{equation}
\mathcal{L}_{\text{CE}}^{PT} = - \mathbb{E}  
\sum_{k \in \Omega}  
\log p_\theta (\bar{z}_i \mid Z_{\mathbf{M}}, T, {P}),
\end{equation}  
To further amplify the influence of pose signals, we extract the pose control signals from the generated dance sequence via Gumbel-Softmax sampling and minimize the discrepancy  $D(P, \hat{P})$ between input pose signals $P$ and those extracted from the output $\hat{P}$, i.e., 
\begin{align}\label{dist}
     D(P, \hat{P})  &=  \frac{ \sum_{i\in \mathcal{N}} \sum_{j \in \mathcal{J}} I(i,j) \| \hat{P}_{i, j} - P_{i, j} \|^2_2 }{\sum_{i\in \mathcal{N}} \sum_{j \in \mathcal{J}} I(i,j)}
\end{align}
where $I(i,j)$ is a binary value indicating whether the pose signal $P$ contains a valid value at frame i for joint j. The total loss function, $\mathcal{L}_{\text{PT}} = \mathcal{L}_{\text{CE}}^{PT} + \lambda_D \mathcal{L}_{D}$, is the weighted sum of CE loss and discrepancy loss $\mathcal{L}_{D} = \mathbb{E}[D(P, \hat{P})]$, 

\noindent \textbf{Progressive Training.}  Integrating music, text, and pose control signals into a dance generation model is challenging due to conflicting loss functions that create competing gradients during training. In particular, the kinematic terms in the music tower loss ($\mathcal{L}_{MT}$) aim to recover the full body skeleton joints, while the discrepancy term in the pose tower loss ($\mathcal{L}_{PT}$) focuses on restoring only the non-editable joints. To address this, we adopt multimodal progressive training, which incrementally incorporates modality-specific signals. This approach allows each modality of the model to be trained according to its respective loss function, mitigating gradient conflicts. The training process begins with training T2M model on text-to-motion dataset using textual prompts that describe a broad range of human actions. Next, we introduce the music tower, trained on dance-only datasets while keeping the T2M model frozen. This stage incorporates both music and genre prompts. At last, pose tower is trained with the frozen T2M using pose and genre prompts. Although the pose tower is not explicitly trained with music signal $A$, integrating the music and pose towers with the T2M allows the pose $P$ and music $A$ signals to simultaneously manipulate the prior motion distribution learned by the T2M. 

\begin{table*}[t]
\centering
\caption{Performance comparison on FineDance dataset. Red bold indicates the best and blue underline the second-best results.}

\scalebox{0.9}{
\begin{tabular}{cccccccc}
\hline
    Methods
    & \multicolumn{2}{c}{Motion Quality $\downarrow$}
    & Foot Skating Ratio $\downarrow$
    & \multicolumn{2}{c}{Motion Diversity $\uparrow$}
    & BAS $\uparrow$
    & Run Time (s) $\downarrow$\\
\cmidrule(r){2-3} \cmidrule(l){4-4} \cmidrule(l){5-6}
 & FID$_k$ & FID$_g$ & FSR & Div$_k$ & Div$_g$ &  &  \\
\hline
FACT \cite{li2021ai} & 113.38 & 97.05 & 28.44\% & 3.36 & 6.37 & 0.1831 & 9.46 \\
MNET \cite{kim2022brand} & 104.71 & 90.31 & 39.36\% & 3.12 & 6.14 & 0.1864 & 10.26 \\
Bailando \cite{siyao2022bailando} &  82.81 & \color{blue}{28.17} & 18.76\% & 7.74 & 6.25 & 0.2029 & \color{blue}{\underline{1.46}} \\
EDGE \cite{tseng2023edge} &  94.34 & 50.38 & 20.04\% & \color{blue}{\underline{8.13}} & \color{blue}{\underline{6.45}} & 0.2116 & 2.27 \\
LODGE \cite{li2024lodge}  & \color{blue}{45.56} & 34.29 & \color{red}{\textbf{5.01 \%}} & 6.75 & 5.64 & \color{red}{\textbf{0.2397}} & 8.16 \\
\hline  
\rowcolor{gray!15}
\textbf{DanceMosaic (ours)} & \color{red}{\textbf{22.33}} & \color{red}{\textbf{12.96}} & \color{blue}{\textbf{\underline{7.58\%}}} & \color{red}{\textbf{8.54}} & \color{red}{\textbf{8.36}} & \color{blue}{\textbf{\underline{0.2190}}} & \color{red}{\textbf{0.8}} \\
\hline
\end{tabular}
}
\label{tab:FineDance}
\end{table*}

\begin{figure}[ht]
    \centering
    \includegraphics[width=0.95\linewidth]{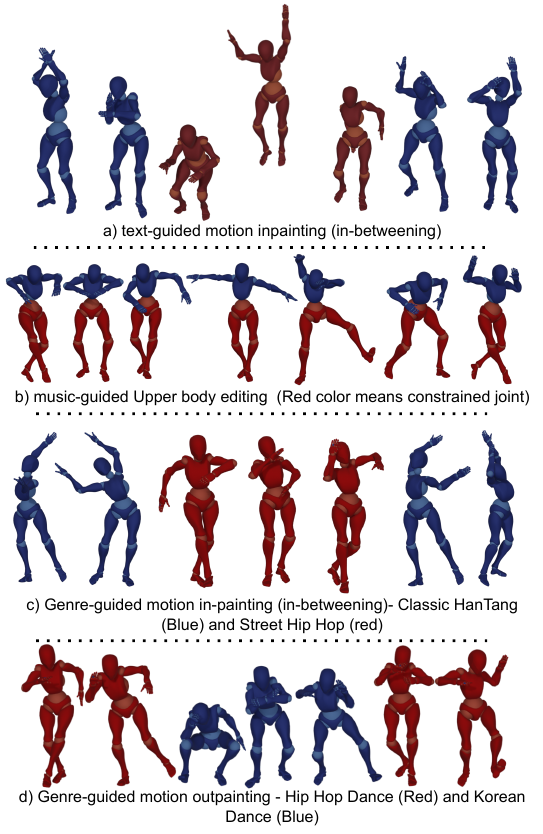}
    \caption{Various Applications Using DanceMosaic}
    \label{fig:applications}
\end{figure}

\subsection{Multimodal Guidance at Inference}
\label{sec:inference_opt}
\textbf{Multimodal Classifier-free Logits Guidance. }
To enhance the influence of music, genre, and pose signals, we adopt Multimodal Classifier-free Logits Guidance, inspired by Classifier-Free Diffusion Guidance \cite{cfg},  within our masked dance model. As shown in Fig. \ref{fig:main}, the final logits from token classifier are computed as a linear combination of unconditional logits and conditional logits informed by text, music, and pose inputs.  These combined logits are transformed through a softmax function to yield the token distribution \( p_\theta (\bar{z}_k \mid Z_{\mathbf{M}}, T, A, C) \).  During inference, motion tokens are iteratively generated using this learned distribution as a decoding confidence measure. Starting with a fully masked sequence of length \( L \), the model performs \( T \)  decoding iterations. At each iteration $t$, masked tokens are sampled from  \( p_\theta (\bar{z}_k \mid Z_{\mathbf{M}}, T, A, C) \). Tokens with low confidence are remasked and resampled, with the number of remasked tokens \(n_M\) following a decaying schedule 
 \( n_M = L \cos\left(\frac{\pi}{2} \frac{t}{T}\right) \)  \cite{chang2022maskgitmaskedgenerativeimage, pinyoanuntapong2024mmm}. This schedule applies a higher masking ratio in early iterations when confidence is low and gradually reduces it as more context is available, shown in Fig. \ref{fig:inferece_dancemosaic}(a).  

 \noindent \textbf{Pose-guided Token Optimization.} The sparse pose control signals have a subtler influence on motion distribution compared to textual and musical cues. Thus, the pose tower alone is not sufficient to accurately incorporate the editing constraints. To enhance motion quality and enforce dance motion alignment with sparse pose guidance, we further refine motion token embeddings $\bar{Z}$ via gradient descent to minimize the discrepancy function in Eq. \eqref{dist}:
\begin{equation}
\bar{Z}^+ = \arg\min_{\bar{Z}} D(P, \hat{P})
\end{equation}
 where motion tokens are iteratively updated as, 
\begin{equation}
\bar{Z}_{t+1} = \bar{Z}_t - \eta \nabla_{\bar{Z}}  D(P, \hat{P}) 
\end{equation}
During inference-time optimization illustrated in Fig. \ref{fig:inferece_dancemosaic}(b), the network remains frozen, with only input motion tokens $\bar{Z}$ being trainable, thus accelerating the optimization process. Optimized input tokens $\bar{Z}^+$ pass through classifier to alter the logits and manipulate the sampling probabilities. 

\subsection{Applications}
\partitle{Spatial Body Part Editing} As shown in Fig. \ref{fig:applications}(b), DanceMosaic enables precise joint-specific control by progressively refining motion sequences through pose tower and inference-time optimization. It first generates a base motion sequence, then iteratively adjusts designated editable joints while preserving global coherence. By integrating music-aware and pose control signals, DanceMosaic achieves fine-grained upper- and lower-body refinements, adapting seamlessly to diverse music styles and text prompts for expressive, rhythmically aligned dance synthesis.

\partitle{Temporal Body Part Editing} As shown in Fig. \ref{fig:applications}(c), DanceMosaic leverages its generative masking to interpolate missing motion segments while ensuring temporal consistency and rhythmic alignment. By strategically placing \texttt{[MASK]} tokens, it reconstructs smooth transitions between keyframes using music and text prompts. Trained on diverse conditional masking patterns, DanceMosaic achieves natural, fluid temporal motion editing without additional supervision, maintaining spatial and musical coherence.

\partitle{Long Dance Generation} DanceMosaic enables zero-shot long-form dance synthesis without the need for retraining. Given a sequence of music and text signals, it first generates individual motion segments. Then, as shown in Fig. \ref{fig:applications}(d), through masked token reconstruction, it synthesizes transitions by conditioning on adjacent sequences, ensuring rhythmically aligned continuity while preserving temporal coherence and expressive flow for extended performances.

\section{Experiments}

\partitle{Datasets}  
We utilized  FineDance \citep{li2023finedance}, a 7.7 h corpus of 202 dance sequences across 16 genres, each paired with high-quality audio and detailed annotations including per-frame music features and genre labels. This rich, multimodal alignment makes FineDance an ideal benchmark for our multi-tower architecture and inference-time editing capabilities. All sequences are down-sampled to 20 fps, and we use the authors’ original train/val/test splits. For pretraining, we leverage HumanML3D \citep{guo2022generating}, a text-driven motion dataset with 14,616 sequences and 44,970 captions, to learn a robust T2M backbone. We train a motion tokenizer on the union of HumanML3D and FineDance. The T2M backbone is pretrained on HumanML3D, while the Music and Pose Towers are trained exclusively on FineDance.  

\begin{table}[t]
\centering
\caption{Comparison of text-conditional motion synthesis on HumanML3D \cite{guo2022generating} test set}
\scalebox{0.65}{
\begin{tabular}{lccccc}
\hline
\textbf{Methods} & \textbf{Top-1 $\uparrow$} & \textbf{FID $\downarrow$} & \textbf{MM Dist $\downarrow$} & \textbf{Diversity $\uparrow$} \\
\hline
Dance2Music \cite{Dancingtomusic} & 0.033 & 66.98 & 8.116 & 0.725 \\
TM2T \cite{TM2T} & 0.424 & 1.501 & 3.467 & 8.589 \\
TM2D  \cite{gong2023tm2dbimodalitydriven3d} & 0.319 & 1.021 & 4.098 & 9.513 \\
MoMask \cite{momask} & \textcolor{blue}{\underline{0.521}} & \textcolor{blue}{\underline{0.045}} & \textcolor{blue}{\underline{2.958}} & \textcolor{blue}{\underline{9.624}} \\
MMM \cite{pinyoanuntapong2024mmm} & 0.515 & 0.089 & \textcolor{red}{\textbf{2.926}} & 9.577 \\
\hline
\rowcolor{gray!15}  \textbf{DanceMosaic (our T2M)} & \textbf{\textcolor{red}{0.523}} & \textcolor{red}{\textbf{0.039}} & 3.042 & \textcolor{red}{\textbf{10.445}} \\
\hline
\end{tabular}
}
\label{tab:motion_comparison}
\end{table}

\partitle{Evaluation Metrics} We evaluate DanceMosaic using kinematic and geometric Fréchet Inception Distances (FID$_k$, FID$_g$) to measure how closely our generated motions match real‐world dynamics and pose geometry, and mean pairwise L$_2$ distances in the same feature spaces (Div$_k$, Div$_g$) to capture motion diversity. We further report Beat Alignment Score (BAS) for music-motion synchronization, Foot Skating Ratio (FSR) for physical plausibility, and average inference time per 9s clip to show real‐time capability. For \textit{\textbf{dance generation evaluation}}, we adopt the evaluation protocol of LODGE \citep{li2024lodge} and all results averaged from 20 runs. All experiments are done on single RTX A5000 GPUs.  

\subsection{Quantitative Comparison to State-of-the-art}

\partitle{Music-Dance Motion Generation} DanceMosaic achieves state-of-the-art results, shown in Table \ref{tab:FineDance}, on FineDance by combining a multi-tower generative masked model with synchronized progressive training and multimodal inference guidance. This design delivers real-time, editable 3D dance synthesis without sacrificing realism or musical alignment. In motion quality, kinematic FID$_k$ falls from 45.56 to 22.33 (-51\%) and geometric FID$_g$ from 34.29 to 12.96 (-62\%), demonstrating a much closer match to real dance distributions. Geometric diversity (Div$_g$) rises from 6.45 to 8.36 (+30\%), while kinematic diversity (Div$_k$) remains high at 8.54, reflecting richer, varied motion. Foot skating ratio (FSR) drops from 18.76\% to 7.58\%, indicating more physically plausible sequences. Beat alignment stays strong (BAS = 0.219), confirming that enhanced realism does not compromise synchronization. Finally, inference runs in 0.80s per 9s clip, over 2× faster than Bailando and 10× faster than LODGE, validating DanceMosaic’s real-time efficiency.


\partitle{Text-to-Motion Generation} To demonstrate generalization beyond music-driven dance generation, we evaluate our Text-to-Motion Backbone (T2M) on HumanML3D~\citep{guo2022generating} (Fig.~\ref{fig:main}(b)), and all results are averaged from 20 runs. Table~\ref{tab:motion_comparison} demonstrates that, with only the T2M component active, DanceMosaic achieves the highest Top-1 accuracy (19.57\% relative gain) and the lowest FID among all baselines. It also delivers strong motion diversity (10.45) and multimodal distance (3.04), striking an excellent balance between realism, variety, and alignment to text prompts. These results confirm the T2M backbone’s capability to generate high-quality motions from natural language descriptions.


\begin{table}
\centering
\caption{Effectiveness of DanceMosaic Components}
\scalebox{0.95}{
\begin{tabular}{cccccc}
\hline
    \multicolumn{3}{c}{{Components}}
    & \multicolumn{2}{c}{{Motion Quality}}
    & \\
\cmidrule(r){1-3}\cmidrule(r){4-5}
Motion Prior & Text & Music & FID$_k$ $\downarrow$ & Div$_k$ $\uparrow$ & {BAS $\uparrow$}\\ 
\hline
$\times$ & $\times$ & \checkmark & 45.56 & 6.75 & 0.2014 \\
$\times$ & \checkmark & \checkmark & 43.07 & 4.92 & 0.2122 \\
\checkmark & \checkmark & $\times$ & 58.66 & 3.31 & 0.193 \\
\checkmark & $\times$ & \checkmark & \underline{34.49} & \underline{7.55} & \underline{0.2104} \\
\checkmark & \checkmark & \checkmark & \textbf{22.33} & \textbf{8.54} & \textbf{0.2190} \\
\hline
\end{tabular}
}
\label{tab:Components}
\end{table}

\begin{table}[t]
\centering
\caption{Impact of Music Guidance Scale}
\scalebox{0.95}{
\begin{tabular}{cccc}
\hline
    \makecell{Music Guidance Scale (\(w_A\))}
    & \makecell{FID$_k$ $\downarrow$}
    & \makecell{Div$_k$ $\uparrow$}
    & \makecell{BAS $\uparrow$}\\ 
\hline
0	& 58.65	& 3.30 & 0.1972 \\
0.2	& 56.23	& 4.15 & 0.2007 \\
0.4	& 43.08	& 5.13 & 0.2048 \\
0.6	& 30.87	& 6.53 & 0.2094 \\
0.8	& 27.98	& 7.74 & 0.2105 \\
\textbf{1}	& \textbf{22.33} & \textbf{8.54} & \textbf{0.2190} \\
2	& 37.46	& 7.20 & 0.2301 \\
\hline
\end{tabular}
}
\label{tab:MGS}
\end{table}

\section{Ablation Study}


\partitle{Effectiveness of Proposed DanceMosaic Components} Our ablation study (Table~\ref{tab:Components}) demonstrates that each core component of DanceMosaic is essential to its overall performance. First, the pretrained Text-to-Motion (T2M) backbone serves as a powerful distribution prior: when we remove this backbone, kinematic FID$_k$ jumps, indicating a severe loss of motion fidelity. Next, text guidance, implemented via the genre-prompt embedding, provides crucial semantic structure. Without it, Div$_k$ superficially rises (8.55 versus 6.75), but the motions become erratic and unstructured, underscoring how genre cues anchor style and coherence. Finally, the Music Tower injects rhythmic control: disabling music guidance yields the worst FID$_k$ (58.66) and collapses beat alignment, confirming that explicit audio conditioning is vital for synchronizing movement to music. These results validate our multi-tower masked architecture and synchronized training as the foundation for high-quality, semantically consistent dance synthesis.

\partitle{Impact of Music Guidance Scale}  Table~\ref{tab:MGS} shows that gradually increasing the music guidance weight $w_A$ (with text weight fixed) drives a consistent drop in kinematic FID$_k$ and a rise in Beat Alignment Score, reaching an optimum at $w_A=1$. Beyond this point, further emphasis on music yields diminishing returns, highlighting that our classifier-free logits guidance, by enabling precise, per-modality weighting, strikes the ideal balance between motion realism and rhythmic fidelity.  

\begin{table}[t]
\centering
\caption{Effect of Token Optimization and Pose Tower}
\scalebox{0.8}{
\begin{tabular}{cccccccc}
\hline
    \makecell{Pose Tower}
    & \makecell{ITTO}
    & \makecell{FID$_k$ $\downarrow$}
    & \makecell{FID$_g$ $\uparrow$}
    & \makecell{BAS $\uparrow$}
    & \makecell{Joint Dist. $\downarrow$}\\ 
\hline
\checkmark & $\times$ & 84.56 & 49.74 & 0.2263 & 0.037 \\
$\times$ & \checkmark & 32.37 & 48.80 & 0.2127 & 0.005 \\
\checkmark & \checkmark &  \textbf{22.89} & \textbf{39.11} & \textbf{0.2277} & \textbf{0.004} \\
\hline
\end{tabular}
}
\label{tab:PGM}
\end{table}

\partitle{Effect of Token Optimization and Pose Tower} Table~\ref{tab:PGM} shows that inference-time token optimization (ITTO) and Pose Tower each contribute uniquely to precise pose editing. ITTO alone cuts kinematic FID$_k$ by 62\% (84.56→32.37) and reduces Joint Distance by 86\% (0.037→0.005), demonstrating its power to enforce sparse pose constraints via per-token refinement. The Pose Tower alone yields modest gains in geometric FID$_g$ (49.74→39.11) and leaves joint errors high, indicating that architecture-level conditioning needs the fine-grained adjustments provided by ITTO. Together, they drive FID$_k$ to 22.89 (-73\%), boost BAS to 0.2277, and minimize Joint Distance to 0.004, confirming that our multi-tower masked architecture plus inference-time token optimization enables exact, semantically coherent edits without compromising overall motion quality or musical alignment.

\section{Conclusion}

In this work, we presented DanceMosaic, a novel method designed to address challenges in automatic, high-fidelity, and editable 3D dance motion generation. Our approach leverages a generative masked text-to-motion model as a distribution prior, facilitating the probabilistic mapping from various guidance signals, music, genre-specific text prompts, and spatial pose constraints, to high-quality dance motions. We developed a multi-tower masked motion architecture that incorporates a text-conditioned backbone with two modality-specific branches, music-guidance tower and pose-guidance tower. By employing synchronized and progressive masked training, we effectively integrated the pre-trained text-to-motion prior while enabling each guidance branch to optimize independently and avoiding training conflicts. To further improve the alignment and realism of generated dances, we proposed classifier-free logits guidance and pose-guided inference-time token optimization techniques. Extensive evaluations confirmed that DanceMosaic achieves state-of-the-art performance, surpassing existing methods. 

\section*{Acknowledgment}

This work is supported by UNC Charlotte Art$\times$Sci grant. 

\bibliography{aaai2026}

\end{document}